\documentclass{llncs}
\usepackage{graphicx}
\usepackage[utf8]{inputenc}

\title{A Characterization of Mobility Management in User-centric Networks \thanks{This work has been developed in the context of project UMM: User-centric Mobility Management, reference PTDC/EEA-TEL/105709/2008, sponsored by Fundação para a Ciência e Tecnologia.}}

\author{Andréa Nascimento\inst{1} \and Rute Sofia\inst{1} \and Tiago Condeixa\inst{2} \and Susana Sargento\inst{2}}

\institute{Informatics Systems and Technologies Research Unit, Lusófona University, Portugal \\
\email{\{andrea.nascimento, rute.sofia\}@ulusofona.pt}
\and Instituto de Telecomunicações, University of Aveiro, Portugal  \\
\email{\{tscondeixa, susana\}@ua.pt}
}

\begin{document}

\maketitle

\begin{abstract}
 Mobility management is a key aspect to consider in future Internet architectures, as these architectures include a highly nomadic end-user which often relies on services provided by multi-access networks. In contrast, today's mobility management solutions were designed having in mind simpler scenarios and requirements from the network and where roaming could often be taken care of with previously established agreements. With a more dynamic behavior in the network, and also with a more prominent role from the end-user, mobility management has to deal with additional requirements derived from new Internet paradigms. To assist in understanding such requirements and also how to deal with them, this paper proposes a starting point to dismantle current mobility management notions. Our contribution is an initial proposal on defining mobility management in concrete functional blocks, their interaction, as well as a potential grouping which later can assist in deriving novel and more flexible mobility management architectures.
\keywords{Wireless networks, mobility management, user-centricity}
\end{abstract}

\section{Introduction}\label{intro}

Internet services and models have been going through a paradigm shift, product of three main factors: i) widespread wireless technologies; ii) increasing variety of user-friendly and multimedia-enabled terminals; iii) wider availability of open-source tools for content generation. Together, these three factors are changing the way that Internet services are delivered and consumed as there is a trend where the end-user has a particular role in controlling content as well as connectivity, based upon cooperation. These spontaneous environments, \textit{user-centric} networks, rely on the notion that Internet users carry or own devices that may be part of the network. Hence the human roaming behavior of each user, be it individually or from an aggregate perspective, directly impacts  the way the network is operated and perceived.

Human movement patterns in these environments may exhibit high variability as they are based on individual users routines and on users interests towards targets (e.g. locations, other users). Hence, mobility management is required to ensure adequate connectivity models and adequate network operation to support end-user expectations towards his/her roaming services. Considering the dynamics of user-centric networks and its self-organizing nature, it is crucial to attempt to develop end-to-end mobility management solutions more flexible than the ones existing today, as user-centric wireless networks are starting to heavily populate Internet fringes.

Currently, the most popular solutions for global mobility management have in common a model where a centralized and static \emph{mobility anchor point} is responsible for keeping some form of association between previous and current identities of a mobile node. In user-centric environments, as explained, there is the  need to better understand the roles that a mobility anchor point can have; the best location for these elements; and efficient ways to select the best anchor point for a mobile node. Moreover, considering that user-centric environments are heavily based on the users interests on being part of the network, and also assuming that the users might also control management functionalities, the period of time a mobility anchor point may or may not be available is highly variable. This poses extra stress on seamless and centralized mobility mechanisms, which have to manage handovers more often. 

The aim of this paper is to provide an initial analysis of aspects that have to be considered when attempting to make end-to-end mobility management schemes more flexible. Our expectations are to contribute to an out-of-the-box notion of mobility management, by splitting mobility management as a whole into concrete functional blocks, and by explaining their impact and how to group such blocks. Our model is based on centralized solutions which, independently of the OSI Layer they tackle, are based in the same principles, roles, as well as similar operational behavior. Such splitting and categorization will give rise, in our opinion, to new mobility management architectures which are user-centric and more flexible.

The paper is organized as follows. In section \ref{relatedwork} we describe related work, explaining the contributions that our work provides. Section \ref{ucentric} provides a few examples on user-centric networking scenarios, including for each a brief mobility characterization. In sections \ref{mcharac} and \ref{deconstruct} we describe our study on mobility management, which is a characterization based on the current needs of this emerging user-centric networks, and in section \ref{conclusion} we conclude this work.

\section{Related Work}\label{relatedwork}

This section provides a brief description on current work related to mobility management proposals which are based on different perspectives than the standardized solutions.

Bolla et al. consider the application of overlays to deal with mobility from a global perspective \cite{Bolla2009}. They provide a distributed mobility management scheme where mobility anchor points may be located within customer premises. The mobility anchor point itself is still a centralizing element as all the signaling goes through this device. Following the same line of thought, in order to deal with personal mobility and session migration Bolla et al.  propose an application layer mobility framework \cite{Bolla2009:2} and the usage of a personal address, \textit{``a network identifier dinamically assigned to a specific user for a specific communication session''}. The framework performs functions of personal mobility, terminal handover, session migration, and media adaptation for interactive multimedia applications. Although the authors are focused on addressing specific aspects of environments involving media, they do not attempt to analyze how to globally make mobility management more flexible.

Sofia et al. \cite{Sofia} propose an approach whose main objective is to separate control and data functionalities from the mobility anchor point into two different elements, in order to provide a more flexible mobility management framework, and to assist in developing non-centralized (e.g. distributed or hierarchical) mobility architectures. However, the authors do not present a proposal on how the communication between those separated elements can be performed, nor an analysis on why such splitting was relevant.

Chan \cite{Chan2010} proposed the splitting of a mobility system into three logical functions: home network prefixes allocation, location management and mobility routing. The approach is based on the \emph{Proxy Mobile IPv6} \cite{pmipv6} extension for \emph{Mobile IPv6} \cite{mipv6}, and it is also proposed the usage of two mobility anchor elements, called \emph{Home Mobility Anchor} and \emph{Visited Mobility Anchor}. The main objective is to provide a system with mobility anchors distributed over different networks.

Having in mind the recent trend of flatter mobile network architectures, \emph{Dynamic Mobility Anchoring} \cite{Bertin2009} \cite{Seite2010} addresses the concept of “flattening” by confining mobility support in the access network, e.g. only confining it to access routers through a specific implementation of the application of Proxy Mobile IP. Following the same line of thought, i.e. IP mobility management in flatter mobile networks, Chan \cite{Chan2011} describes the differences between centralized and distributed mobility management systems, as well as a list of potential problems and limitations of a centralized approach when compared with a distributed one.

Condeixa et al. \cite{Condeixa2010} analyzed mobility management assumptions and requirements in user-centric scenarios, debating on challenges that need to be addressed to obtain a global mobility management solution considering user-centricity. The authors point out three major concerns for a mobility management system: binding definition, binding maintenance, and forwarding data problem.

Our work has in common with these approaches the motivation that by splitting, de-centralizing, or decoupling mobility management functionality into different blocks may assist in better understanding how and where to manage mobility. As described, most of today’s attempts of flattening mobility management are being applied in the evolved packet core being the sole reason the urgent need to simplify mobility management. We believe that understanding on how such mechanism may work is key to give rise to new research and business opportunities.

\section{User-centric Networking Notions}\label{ucentric}

User-centric networks are environments where an Internet end-user owns and often carries devices that can share Internet access. These environments and the amount of end-user devices sharing Internet access are expected to grow, despite the limitations imposed by traditional operator-driven Internet communication models. 

In our study, mobility management aspects are addressed from an end-to-end perspective but the analysis is applied in user-centric spontaneous wireless environments, which today correspond to the majority of technical scenarios on the last hop towards the end-user. Our user-centric environments are located within the customer premises region (where residential households, and enterprise environments reside). While in contrast, today's mobility management relies on functional blocks that are on the access or service regions. 

Out of the several possible user-centric scenarios, we consider here three: a regular hotspot, a user-provided network (UPN) and a delay tolerant network (DTN). Each scenario is described both from an architectural perspective, as well as from a mobility characterization perspective.
The line of thought driving this analysis is that these representative scenarios hold different requirements and are based on specific mobility assumptions. Hence, after providing a mobility characterization for each of the scenarios, the section concludes with a discussion which shall result in the identification of mobility functionality blocks, based on common requirements that each of these scenarios attain. A more complete description of these and of additional user-centric scenarios can be found in \cite{d1}:
\begin{itemize}
\item \textbf{Hotspot}: a hotspot scenario corresponds to the regular infrastructure mode in Wireless Fidelity (Wi-Fi) environments. This is currently the most common wireless architecture being deployed around us: each Internet enabled household corresponds to one hotspot. In this scenario mobility of users is local and confined to small regions, e.g. a room, an apartment, a small office. Moreover, if the user moves across different \emph{Access Points (APs)}, then connectivity is expected to be intermittent. In a generic hotspot scenario users’ mobility speed is low (pedestrian). Mobility inside each hotspot scenario is mostly managed at OSI layer 2; however, the IP address of the active user equipment’s interface can change after a break. A key aspect to consider is that if  current mobility management solutions are applied to this scenario, despite the fact that most of the movement is local, the mobility anchor point is located on the access or service regions.
\item \textbf{User-provided networks}: UPNs \cite{Sofia2008} have been applied as complement to existing access networks: they allow expansion of infrastructures across one wireless hop. There is usually one individual or entity (the \textit{Micro-Provider, MP}) which is responsible for sharing his/her connection with N-1 other users (out of a universe of N users, who today belong to a single community). Moreover, a user is, in a specific community, simply identified by a virtual identifier (usually, a set of credentials username and password) which is stored by a \textit{Virtual Operator (VO)} and relied upon whenever the user decides to access the Internet by means of a specific community hotspot.
In these emerging architectures, the nodes that integrate the network are in fact end-user devices which may have additional storage capability and sustain networking services. Such nodes, being carried by end-users, exhibit a highly dynamic behavior. Nodes move frequently following social patterns and based on their carriers interests. The network is also expected to frequently change (and even to experience frequent partitions) due to the fact that such nodes, being portable, are limited in terms of energy resources.
\item \textbf{Delay tolerant networks}: The DTN scenario relates to the need to establish on-the-fly an autonomous network within a disaster region (e.g. after an earthquake) based upon the devices that users in the region control and carry. Hence, such DTN consists of a network composed by users with a common objective (a community), grouped in regions. Some nodes move from region to region, establishing the communication between them (since gateways are mobile). Considering the main purpose of this kind of network, and the specific type of scenario where it is deployed, it is possible to establish behavior patterns on the mobility of the nodes, making possible to predict their location in a given instant and to schedule the delivery of information. In this case, the mobility pattern may also impact the routing process. Users moving may be good candidates to act as gateways, because they present a higher possibility of reaching other regions. It is important to notice that a region may be composed by only one user.
\end{itemize}

Table \ref{tablescen} summarizes the main characteristics related to the scenarios described, concerning inherent characteristics, and mobility behavior of the users on each of the scenarios presented. Based on a detailed analysis of the scenarios described \cite{d1} we consider  a set of parameters that should be taken into account when characterizing any mobility management scheme: \emph{i) identification}, which stands for the device identification both from a user and an access perspective; ii) \emph{network scope}, which relates to the reach of the network; iii) \emph{access control}, which relates to the location of the access control mechanism that is normally applied in each scenario; iv) \emph{movement patterns}, related to the pattern that nodes are expected to exhibit in each scenario when roaming; v)\emph{ pause time behavior}, related to the time that a node exhibits a speed that is zero or close to zero; vi) \emph{handover frequency}, related to the node having to switch between different networks or attachment points; vii) \emph{connectivity sharing}, related to the sharing of Internet access. 

In table \ref{tablescen} we provide a brief analysis on how each of the mentioned parameters relate to the three scenarios described. UPNs stand for a relevant case to address in terms of mobility management, as this scenario exhibits features that are not available on the hotspot scenario. The same conclusion can be drawn by looking at the DTN characterization. Both UPNs and DTNs exhibit aspects that were not considered when devising the current (centralized) mobility solutions.

\begin{table}[h]
\centering
\caption{Summary of mobility characterization across user-centric scenarios.}
\label{tablescen}
\scriptsize
\begin{tabular}{|l||l|l|l|l|l|l|l|}
 \hline
 Scenario/	& Hotspot & UPN & DTN \\
 Parameters	&	&	& \\
 \hline \hline
 Identification	& MAC address,	& Trust 	& Tokens or \\
		& credentials	& management 	& certificates; \\
		& managed by	& scheme 	& public/private \\
		& WISP		& community 	& key pair \\
		&		& credentials 	& \\
 \hline
 Network scope	& Small		& Small-large, 	& Small-large	\\
		& environment,	& e.g. household & but static	\\
		& e.g. household& to village/city;& does not exhibit\\
		& shops,	& varies 	& a quick growth\\
		& universities	& dinamically 	&	\\
 \hline
 Access control	& Centralized,	& Decentralized & Decentralized	\\
		& on the 	& and 		& 		\\
		& provider	& spontaneous	& 		\\
		& 		& 		& 		\\
 \hline
 Node speed	& Low		& High		& Varying	\\
 \hline
 Expected	& Low		& High and	& Low and	\\
 movement	& 		& global	& routine based	\\
 frequency	& 		& 		& 		\\
 \hline
 Mobility 	& Local 	& Human/social 	& Local		\\
 pattern	& mobility;	& patterns; 	& mobility	\\
		& preferred	& short distance& social	\\
		& locations	& traveling	& patterns	\\
		& 		& preferred	&		\\
 \hline
 Pause time	& Long pause	& Mix, depends  & Long		\\
		& times 	& on location	& 		\\
		& 		& and user	& 		\\
		& 		& routine	& 		\\
 \hline
 Handover	& Low		& High		& High	\\
 frequency	& 		& 		&	\\
 \hline
 Connectivity	& None		& Yes		& Yes	\\
 \hline
\end{tabular}
\end{table}

\section{Defining Mobility Management: A Characterization}
\label{mcharac}

This section is dedicated to a proposal on a global architectural definition of mobility management functional blocks, as well as roles based on the scenarios previously described.

\subsection{Elements and Roles}

In a mobility management system, three elements are considered in related literature: the \emph{Mobile Node (MN)}, an end-user device for which a mobility service is provided; a \emph{;Mobility Anchor Point (MAP)}, the element responsible for providing the mobility management service, it may reside in the network (e.g. router or access element) or in a server; and the \emph{Correspondent Node (CN)}, that is any element engaged in active communication with the MN. These are generic roles that are today present in different management solutions, independently of the OSI Layer where the solution resides. For instance,
in MIP \cite{mipv6} the MAP is the Home Agent (HA). In the \emph{Session Initiation Protocol (SIP)} \cite{sip} the MAP is the SIP server. In a 3GPP architecture, the mobility anchor is centralized and located in the core network, having all traffic flowing through it, even if services to be used are locally placed closer to the MN.

Towards the idea of making mobility management more flexible (being the aim a reduced operational cost) Seite et al. and Chan et al. suggest to position the mobility anchors closer to the mobile nodes \cite{Seite2010}, ideally in the first element visible on the path from a MN perspective \cite{Chan2011}. Sofia et al. proposed the separation of management functionalities into two elements, attempting to decouple data plane and control plane \cite{Sofia}. In the proposed
architecture, the HAC (control plane element) is located in a server, and HADs (data plane elements) are positioned in the access nodes, close to mobile nodes. Chan relies on the Proxy Mobile IP \cite{pmipv6}, and also splits the mobility anchor functionalities into three logical blocks \cite{Chan2011}. Although the author states that those functionalities are placed in the home network, they do not need to be placed in the same physical entity. Those works can be considered as a first step towards an architecture where the management functionalities are splitted and distributed in different places in the network.

Such approaches, the positioning of the MAP as well as the definition of interactions between the different roles of mobility management have been object of heavy analysis. Still, today there is not truly consensus in where MAP and additional functionality should reside. Such positioning depends on the network architecture and requirements; on the OSI Layer being tackled, as well as on the overall complexity from a technical and policing perspective. Considering that user-centric networks present particular characteristics (e.g. there is no clear splitting between network elements and end-devices), the current centralized standards may not be suitable. Thus, a novel mobility management approach should be designed for such networks, considering all its particularities and following this trend of rethinking the mobility anchor point element.

Therefore, thinking about mobility management functioning in a fine-grained way, we have identified a group of functionality blocks. Based on the dynamics of  user-centric networks, the first step towards a more suitable mobility management approach is by understanding and further analyzing the basic tasks a mobility management should provide.

\subsection{Functional Blocks}

In order to perform a mobility management characterization, as result of an initial analysis on current available mobility management approaches and standards, we have identified the following mobility management functional blocks:

\begin{itemize}

 \item \textbf{Device identification:} corresponds to the network identification for the MN. Usually the main mechanism for a location management is the association between the device's \emph{known-address} and the device's \emph{real-address}. In MIP, known-address and real-address are IP addresses; in SIP, the known-address is a URI, and the real-address is an IP address. In MIP the device identification control is the Home Agent (HA)/Correspondent Node (CN) cache binding. In SIP, it is the user database used by the Proxy server.

 \item \textbf{Identification database control:} corresponds to the mechanism that is applied to control the database identification. This is normally a block relevant from an access perspective, which today follows a centralized approach.

 \item \textbf{Binding mechanism:} it is the signaling related to the device's register to the mobility system. It creates/updates a record in the identification database control, associating the known-address to the real-address. In MIP it is the Binding Update message sent to a HA/CN. In SIP it is the REGISTER message sent to the Registrar server.

 \item \textbf{Routing or forwarding:} it is the process of intercepting the packets destined to the known-address, encapsulating them with the real-address, and forwarding them. In MIP this is performed by the HA; in SIP this process is performed by an element named RTP translator (when it is used).

 \item \textbf{Handover negotiation:} the process taken when the device has its real-address changed. It involves negotiation and signaling. The main objective is to guarantee that the user will keep active all its sessions during the handover process. In MIP, the handover negotiation may be anticipated with the Fast Handover extension \cite{fmipv6}, and the SIP does not implement any anticipation, performing a re-negotiation after the connection between the peers is lost.

 \item \textbf{Resource management:} the resource management is a necessary procedure for the mobility management to guarantee the quality of the connection when the MN changes its point of attachment to the network. However, it is not provided by most of the mobility management approaches. The 802.21 Media Independent Handover (MIH) \cite{mih} standard is focused on the handover process based on a resource management aware negotiation for vertical handovers.

 \item \textbf{Mobility estimation:} it is the procedure of changing the MN point of attachment to the network before its current connection breaks. The extension Fast Handovers for MIP, and the 802.21 MIH provide this functionality.

 \item \textbf{Security/privacy:} it refers to any security or privacy mechanism used to assure the integrity of the elements and signaling in the mobility management system.
\end{itemize}

\subsection{Discussion on Mobility Characterization}

Based on the block characterization there are a few aspects worth to highlight. Firstly, today’s mobility management solutions completely ignore the need for adequate resource management. However, this is a crucial aspect for cellular or wireless networks, in particular for session continuity. Database control is normally centralized, an aspect which may not be compatible with the notion of communities that user-centric networks embody. Routing and forwarding is also
based on mechanisms (e.g. proxy mechanisms) which may not be completely compatible with the fact that users in our scenarios are expected to roam frequently. This is an aspect that can be improved by integrating mobility estimation mechanisms. Security and privacy aspects are also often disregarded.

Moreover, analyzing the identified blocks, one can notice that there are a few categories onto which they seem to be naturally grouped. Firstly, they can be grouped into \textit{data plane} and/or \textit{control plane}. It is also possible to group the functionality blocks into \textit{location management} and/or \textit{handover management} procedures. 

These are aspects that we debate on the next section in an attempt to raise awareness to new and more flexible mobility management schemes.

\section{Deconstructing Mobility Management Centralized Approaches}\label{deconstruct}

This section delves into the potential development of a mobility management architecture that is more adequate to the emerging wireless scenarios described in section 3.

As of today the functional blocks described reside both on the MN and mobility anchor point, being the functionality fully controlled in the later one, which is physically located in the access or service regions. Our aim in analyzing initial forms of deconstructing the need for a centralized mobility management scheme is motivated by the need to find simple and operational ways to split such functionality, as well as ways to “push” such functionality closer to the end-user, having in mind an optimization of mobility management in the context of the scenarios described.

\subsection{Location and Handover Management Categorization}

Mobility management usually is mentioned as consisting of two main blocks: location management and handover management. Location management is the block responsible for locating the devices, i.e. for guaranteeing that they are always reachable, independent of their point of attachment to the network. The handover management block is responsible for maintaining active sessions while MNs roam. Therefore, from a high level perspective, mobility management functionality can be split into these two main blocks. Today, these blocks both reside on the mobility anchor point and are based on information provided by the MN. Solutions such as the \emph{Host Identity Protocol (HIP)} \cite{hip} attempt to provide a decoupling by isolating location management and handover management. Other solutions (e.g. Hierarchical Mobile IP \cite{hmipv6}) optimize handover management by scoping the extent of the impact of such negotiation. 

\subsection{Control and Data Plane Categorization}

\begin{figure}[h]
\centering
\includegraphics[width=3.8in]{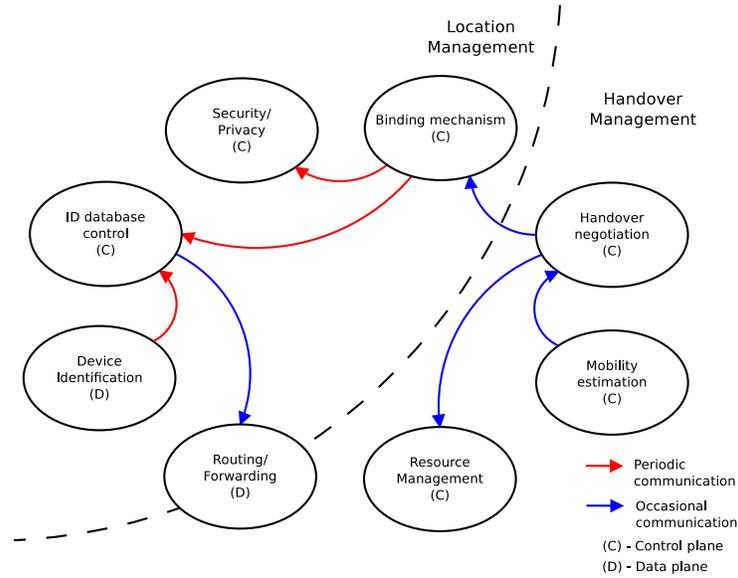}
\caption{Mobility management functional blocks.}
\label{esquema}
\end{figure}

Another way to categorize mobility management functionality is to consider a splitting between control and data planes. As part of the control plane we can cite all the procedures related to the signaling, and the data plane is related to the data traffic, routing, forwarding and address translation. Figure \ref{esquema} shows the relationship between the blocks, in order to identify the communication between them. It shows also the classification concerning data and control planes, and location and handover management.

Between the functional blocks, it is possible to identify two types of communication, in regards to its periodicity. \textit{Periodic communication} is related to procedures that need to be performed in a regular basis, in order to maintain the system updated. The \textit{occasional communication} is related to the procedures performed only as result of a change in the system, for instance, when a MN performs a handover from one point of attachment to another.

Usually, all the communication between the blocks of the handover management side of the picture is triggered when a node movement is detected, or predicted. When a handover is detected, the mobility estimation block triggers the handover negotiation, which will take part in the process. The handover negotiation needs to consult the resource management in order to guarantee that the user will be “always best connected”. For the handover process to complete, the binding mechanism is triggered, so it can update the location information in the identification database control. The identification database control then updates the information in the element responsible for routing/forwarding.

The binding mechanism has a periodic communication with the ID database control, because it is the procedure performed to maintain the ID database control updated. It needs to use the security/privacy procedures to guarantee that no third part could take place in the communication.

\subsection{User Perspective and Access Perspective Categorization}

Currently, the available mobility management approaches offer most of the functionalities described here, but none of those approaches offer all of the functionalities. Those functionalities are placed in different locations in the network and customer premises, and most of them are centralized in one unique element (usually the mobility anchor point). By taking this perspective, we can categorize the blocks into two groups, blocks located in the user perspective and in the access perspective as provided in table \ref{table2}.

\begin{table}
\centering
\caption{Location of mobility management functional blocks.}
\label{table2}
\begin{tabular}{|l||l|}
 \hline
 Parameter & Access and user perspective \\ 
           & categorization \\
 \hline \hline
 Device Identification & User \\
 \hline
 ID database control & Access \\
 \hline
 Binding mechanism & User and access \\
 \hline
 Routing / Forwarding & Access \\
 \hline
 Handover negotiation & User  \\
 \hline
 Resource management & Access and user \\
 \hline
 Security/privacy  & User \\
 \hline
 Mobility estimation & Access and user \\
 \hline
\end{tabular}
\end{table}

Table \ref{table2} shows the current location of each block. It is important to notice that this location is based on current mobility management approaches functioning.

\section{Conclusion}\label{conclusion}

This paper provides a study and a new perspective on ways to make end-to-end mobility management schemes more flexible, being the motivation the fact that user-centricity and in particular user-centric environments are a crucial part of the future of the Internet. We went over three different cases of spontaneous wireless deployments abounding around us, and characterized each from a mobility perspective. Based on such characterization we have derived a set of parameters and functional blocks, and discussed ways to attempt to de-construct the need for centralized architectures, starting by proposing concrete categories to tackle.

As follow-up of this work we intend to take advantage on the blocks identification and data/control planes and location/handover management categorizations to evaluate what is the best location for each of the identified functional blocks. Focusing on the user-centricity, the objective is to perform a deeper study on each of those functionality blocks, in order to identify which of them could be placed into customer premises equipment. Placing mobility management functionalities in the customer premises could provide a mobility system user-centric and independent of the access network. A deeper study should clarify if that is possible, and what is the cost to maintain such approach. Hence, as next steps we intend to address ways to bring mobility management closer to the customer premises in a way that is adequate for the network, while keeping the end-user agnostic in regards to the complexity. A second step to be considered is to analyze such splitting based on the potential impact that it may have both from an end-user and from an access perspective. 

\bibliographystyle{splncs03}

\end{document}